\documentclass[twocolumn,prl,superscriptaddress,showpacs,preprintnumbers,amsmath,amssymb]{revtex4}

\usepackage{graphicx}% Include figure files
\usepackage{dcolumn}% Align table columns on decimal point
\usepackage{bm}% bold math
\usepackage{color}

\newcommand{\gammadot}[0]{\dot{\gamma}}
\newcommand{\Pe}{P_{\rm e}}
\newcommand{\odif}[2]{\frac{\partial #1}{\partial #2}}

\newcommand{\ave}[1]{\langle {#1} \rangle}

\renewcommand{\phi}{\varphi}
\newcommand{\be}{\begin{equation}}
\newcommand{\ee}{\end{equation}}

\begin{document}
\title{Unified study of glass and jamming rheology in soft 
particle systems}

\author{Atsushi Ikeda}
\affiliation{Laboratoire Charles Coulomb,
UMR 5221 CNRS and Universit\'e Montpellier 2, Montpellier, France}

\author{Ludovic Berthier}
\affiliation{Laboratoire Charles Coulomb,
UMR 5221 CNRS and Universit\'e Montpellier 2, Montpellier, France}

\author{Peter Sollich}
\affiliation{King's College London, Department of Mathematics, 
Strand, London WC2R 2LS, United Kingdom}

\date{\today}
\begin{abstract}
We explore numerically the shear rheology of
soft repulsive particles at large volume fraction. 
The interplay between viscous dissipation and thermal motion results in
multiple rheological regimes 
encompassing Newtonian, shear-thinning and yield stress regimes 
near the `colloidal' glass transition when thermal fluctuations 
are important, crossing over 
to qualitatively similar regimes near the `jamming' transition 
when dissipation dominates. In the crossover regime,
glass and jamming sectors coexist and give complex flow curves.  
Although glass and jamming limits are characterized
by similar macroscopic flow curves, we show that they occur over
distinct time and stress scales and correspond to 
distinct microscopic dynamics.
We propose a simple rheological model describing the 
glass to jamming crossover in the flow curves, and discuss the experimental 
implications of our results. 
\end{abstract}


% 62.20.-x - Mechanical properties of solids
% 83.80.Iz - Emulsions and foams
% 83.60.La - Viscoplasticity; yield stress 

\maketitle

The emergence of solidity in disordered assemblies of repulsive particles 
is a well-known phenomenon~\cite{coussot} which remains 
difficult to understand at a fundamental level~\cite{rmp,jammingrev}. 
When compressed, a colloidal suspension undergoes 
a `glass transition' from (metastable) thermal equilibrium~\cite{pusey}, 
as observed experimentally for a broad 
spectrum of particle types~\cite{crocker}. For colloidal hard spheres suspended
in a solvent of viscosity $\eta_s$, the shear viscosity, $\eta_T$,
is a universal function of the packing fraction $\phi$, 
$\eta_T / \eta_s = G(\phi)$, independently of e.g.\
particle size~\cite{chaikin}. 
Solidity also emerges far from equilibrium 
upon compressing non-Brownian 
suspensions of repulsive particles across the
`jamming transition', as in foams or granular 
materials~\cite{jammingrev,crocker}.
The viscosity $\eta_0$ 
of a non-Brownian hard sphere 
suspension is again universal, 
$\eta_0 / \eta_s = J(\phi)$~\cite{lindner}.
Depending on the community and the particular system at hand, 
rheologists use a broad variety of 
functional forms and empirical models 
for $G(\phi)$ and $J(\phi)$, while underlying 
physical processes for both limits are often not distinguished 
in rheology textbooks~\cite{larson}. Our aim is to determine if and 
how these two ideal limits 
are interrelated, addressing also the non-linear 
rheological regimes and the additional effects of 
particle size and softness.

Glass and jamming transitions share
important similarities, in particular at the rheological level. 
In both cases, solidity emerges near a `critical' volume fraction below which 
the material is a fluid whose viscosity increases rapidly
with $\phi$. The amorphous
solid at large density responds elastically for small 
deformation, but flows when a stress larger than a yield stress is 
applied~\cite{coussot}. The 
dynamics becomes very heterogeneous near the critical density; 
its spatial correlations are
usually interpreted by appealing to underlying 
phase transitions~\cite{book},
though the nature of these remains a subject of debate~\cite{rmp,jammingrev}. 
Based on these similarities, a unified jamming phase diagram has been proposed
where thermal and athermal systems appear as a 
single `jammed' phase~\cite{liunagel,pica}.

At the theoretical level, recent results have clarified the 
relation between glass and jammed phases~\cite{jorge,PZ,hugo}, 
suggesting that the jamming transition occurs well inside 
the non-ergodic glassy phase. For systems of soft 
repulsive particles, as studied below, a glass transition line 
$T_G(\phi)$ separates the fluid (at high $T$, low $\phi$)
from the glass (at low $T$, large $\phi$), while the jamming 
transition occurs upon compression at $T=0$ 
inside the glass phase~\cite{hugo}. 
However, these static calculations shed little light on either
dynamical properties or rheology~\cite{thomas}. 
Although this theoretical scenario appears 
in broad agreement with numerical work~\cite{pica,tom}, 
glass and jamming transitions are typically located using different
sets of methods and observables, normally requiring
extrapolation~\cite{tom}.
Similar ambiguities exist in experimental work where
e.g.\ estimates for the location of the colloidal glass 
transition cover the range $\phi_G \approx 0.57 \ldots 0.635$, 
depending on how the divergence of $G(\phi)$ is 
extrapolated~\cite{pusey,chaikin,gio}.
In the same vein, data for the divergence of $J(\phi)$ for athermal suspensions
lie in the range $\phi_J \approx 
0.585 \ldots 0.66$~\cite{lindner,tom,pouliquen}.

In this paper, we argue that a clearer picture emerges when 
the non-linear rheology of both thermal and athermal suspensions 
is considered. We use computer simulations to investigate the 
flow properties of concentrated assemblies of soft 
repulsive particles, and vary the relative strength 
of thermal fluctuations and viscous dissipation to study 
the crossover from thermalized suspensions
(relevant to soft colloids) to purely athermal ones 
(relevant to jammed solids) within a single computational 
framework. This setting allows us to observe 
and disentangle multiple rheological regimes within a single system,
establishing in particular unambiguously that the increases 
of the shear viscosities $\eta_T(\phi)$ and $\eta_0(\phi)$ upon compression 
are unrelated. This has important consequences
for the jamming phase diagram of soft particles.
Although glass and jamming limits are characterized
by similar macroscopic flow curves, we also show that 
they in fact occur over well-separated time and stress 
scales and correspond to qualitatively
different microscopic dynamics.

We analyze theoretically the behaviour of  
sheared assemblies of soft repulsive particles 
immersed in a solvent, such as star polymers, microgels, or 
dense emulsions~\cite{crocker}. 
The simplest way to model these systems at large packing fraction
is to ignore hydrodynamic interactions 
and consider only pairwise repulsion between particles, 
such as $V(r) = \epsilon (1-r/a)^\alpha 
\Theta(a - r)$, where $\Theta(x)$ is the Heaviside function, and $a$ is 
the particle diameter~\cite{durian}.
We use molecular dynamics simulations to study the 
steady state rheology of harmonic spheres, $\alpha=2$,
in a simple shear flow. We simulate the following Langevin dynamics, 
\begin{eqnarray}
\xi \left(\odif{\vec{r}_i}{t} - \gammadot y_i \vec{e}_x\right) 
= - \sum_{j \neq i}  
\odif{V(|\vec{r}_i - \vec{r}_j|)}{\vec{r}_i} + \vec{f}_i(t), 
\label{langevin}
\end{eqnarray}
where $\vec{r}_i$ and $y_i$ represent the position 
and the $y$-coordinate of particle $i$, respectively, 
and $\vec{e}_x$ is the unit vector along the $x$-axis. 
The damping coefficient, $\xi$, which accounts for 
viscous dissipation,
and the random force $\vec{f}_i$ describing thermal fluctuations 
obey the fluctuation-dissipation relation, 
$\ave{\vec{f}_{i}(t) \vec{f}_{j}(t')^{\rm T}} = 2 k_B T \xi \delta_{ij} 
{\bf 1} \delta(t-t')$, where $k_B$ is Boltzmann's constant.

The Langevin dynamics in Eq.~(\ref{langevin}) is characterized by
two microscopic timescales: $\tau_0 = \xi a^2 / \epsilon$
controls the dissipation, 
while $\tau_D = \xi a^2 / (k_B T) = \tau_0 \epsilon / (k_B T)$
sets the timescale for Brownian motion. 
Therefore, $\tau_D$ and $\tau_0$ are comparable at high 
temperatures but become well-separated when $k_B T \ll \epsilon$, 
with $\tau_D \gg \tau_0$. The shear rate $\gammadot$ introduces 
a third timescale, $\gammadot^{-1}$, from which  
the P\'eclet number is defined, $\Pe = \gammadot \tau_D$. 
Timescale separation 
at low $T$ allows us to separately explore the thermal regime 
at small $\gammadot$, $\Pe \ll 1$, where Brownian motion is relevant, 
and the athermal regime at larger $\gammadot$, $\Pe \gg 1$, 
where the suspension is non-Brownian. In the alternative 
SLLOD dynamics frequently used to shear
suspensions~\cite{allen}, inertia is included and 
the thermostat is the only source of dissipation. Thus,
the only accessible regime is $\Pe <1$
(with now $\tau_D = \sqrt{\frac{m a^2}{k_BT}}$),
and the $T \to 0$ limit is unphysical.
By contrast, Eq.~(\ref{langevin}) can be simulated
at $T=0$ (i.e., $\Pe = \infty$), where the dynamics 
becomes similar to earlier studies of the jamming 
transition~\cite{olson}. Thus, Eq.~(\ref{langevin})
allows us to study thermal and athermal systems under shear 
in a unified manner. 

We study a $3d$ system of $N=10^3$ harmonic spheres, using 
a 50:50 mixture of spheres with diameter ratio 1.4
to avoid crystallisation~\cite{olson,tom}. We measure  
length in units of the small particle diameter, $a$, 
time in units of $\tau_0$ and temperature in units of $\epsilon/k_B$.
We integrate Eq.~(\ref{langevin}) at constant $\gammadot$ 
using Heun's method 
with Lees-Edwards periodic boundary conditions~\cite{allen}
over a typical simulation time of at least $10/\gammadot$. 
We evaluate the $xy$-component of the shear stress, $\sigma$, using the 
Irving-Kirkwood formula, and deduce the shear viscosity,
$\eta = \sigma / \gammadot$. 
The stress and viscosity units are respectively $\epsilon/a^3$ and $\xi/a$. 
For thermal simulations at low temperatures, we combine data from Langevin and
SLLOD dynamics to broaden the range of shear rates towards 
very low P\'eclet numbers, where SLLOD is more efficient.
We have checked that both methods yield equivalent results 
at equal $\Pe$ values~\footnote{Some simulations 
at very low $\phi$ and large $\gammadot$ 
yield configurations with ordering along the shear flow. We discard 
those state points.}.   

\begin{figure}
\includegraphics[width=1.0\columnwidth]{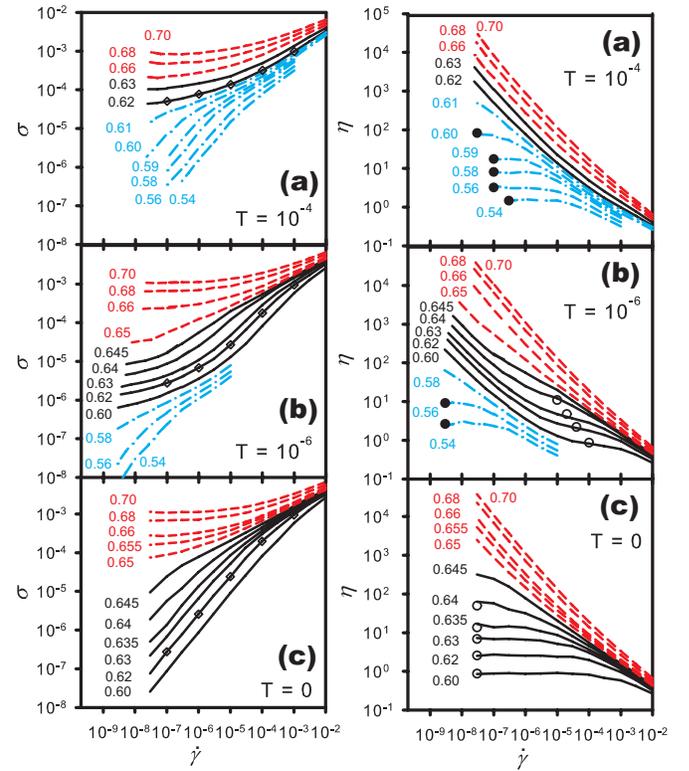}
\caption{Flow curves at different temperatures and densities, 
shown as $\sigma(\gammadot)$ (left) or $\eta(\gammadot)$ (right).
Diamonds (left) mark the state points 
analyzed in Fig.~\ref{msd}. Circles (right)
indicate thermal (closed) and athermal (open) viscosities
reported in Fig.~\ref{zeroshear}. 
Flow curves are shown in blue (dot-dashed) when thermal Newtonian
behaviour is observed, in red (dashed) when thermal effects are negligible,
in black otherwise.}
\label{flow}
\end{figure}
 
\begin{figure}
\begin{center}
\includegraphics[width=1.0\columnwidth]{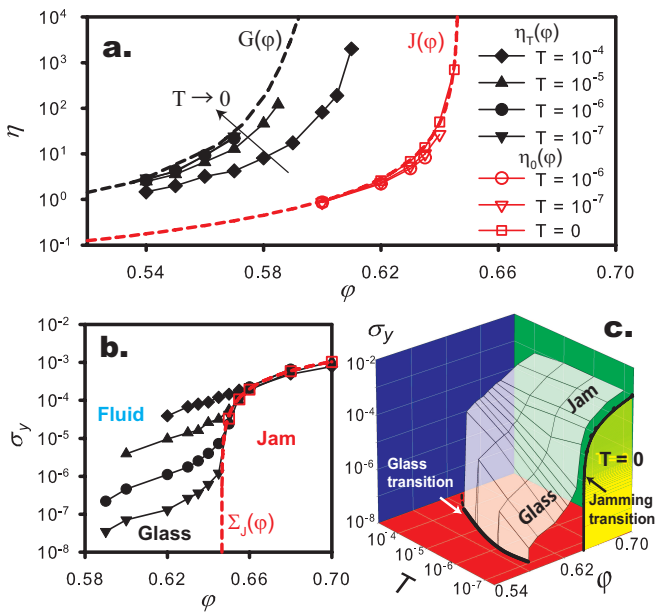}
\end{center}
\caption{(a) Shear viscosities $\eta_T$ and $\eta_0$
and their distinct hard sphere limits $G(\phi)$~\cite{gio}
and $J(\phi)$~\cite{olson}.
(b) Density dependence of yield stress 
for different temperatures, including the $T=0$
limit. 
(c) Same data as in (b) in a `glass-jamming phase diagram'.}
\label{zeroshear}
\end{figure}

We first characterize the macroscopic flow properties 
at $T=10^{-4}$. Since
$\tau_D = 10^4$, the data in Fig.~\ref{flow}a mainly cover the
thermal range, $\Pe < 1$.  Accordingly, the resulting flow curves are typical 
of dense fluids sheared across the glass 
transition~\cite{thomas,yamamoto,BB}. Briefly, for densities 
$\phi < \phi_G \approx 0.61$, 
flow curves are Newtonian at low shear rates, while for 
larger $\gammadot$, the external flow accelerates structural 
relaxation leading to shear-thinning. 
The Newtonian viscosity, $\eta_T (\phi)$, increases rapidly upon 
compression towards $\phi_G$. Above $\phi_G$, 
the linear viscosity is too large to be measured, and the 
system behaves instead as a solid (a glass) with a finite yield stress, 
defined as $\sigma_Y = \lim_{\gammadot \to 0} \sigma(\gammadot)$. 
Both the shear viscosity $\eta_T(\phi)$ and the yield stress $\sigma_Y(\phi)$ 
can be used to locate the glass transition, 
which corresponds to either the divergence of $\eta_T$, or the 
emergence of a finite $\sigma_Y$, see Fig.~\ref{zeroshear}.

At significantly lower temperature, 
$T = 10^{-6}$, the above phenomenology persists as long 
as $\Pe < 1$, which now corresponds to 
very low shear rates, $\gammadot < \tau_D^{-1} = 10^{-6}$ in 
Fig.~\ref{flow}b. Thus, we can determine a 
Newtonian viscosity $\eta_T(\phi)$ for $\gammadot \to 0$
and $\phi < \phi_G \approx 0.59$, and a finite yield stress above $\phi_G$. 
Note that $\phi_G$ decreases slowly with decreasing $T$, see 
Ref.~\cite{tom}.  
Because $\tau_D$ is very large,  
there now exists a broad $\gammadot$ window where $\Pe \gg 1$
and thermal fluctuations play little role. Surprisingly, the data in 
Fig.~\ref{flow}b show that for a range of 
densities {\it above} the glass transition, $0.59 < \phi < 0.63$, 
the system flows as a simple Newtonian fluid when $\Pe \gg 1$.
This allows us to define a second viscosity, $\eta_0$, that
grows upon compression towards $\phi_J \approx 0.64$. 
Finally, for $\phi > \phi_J$, the flow curves
are mainly characterized by
a yield stress, $\sigma_Y(\phi)$.
The shear viscosities 
$\eta_T(\phi)$ and $\eta_0(\phi)$
(Fig.~\ref{zeroshear}a) obey
clearly distinct behaviours. While the growth 
of $\eta_T$ reflects the approach to the glass transition, 
$\phi_G \approx 0.59$, the 
increase in $\eta_0$ is separately controlled 
by the jamming transition, $\phi_J \approx 0.64$. 
Given that both viscosities are defined over distinct 
density and shear rate regimes,
and can be simultaneously observed at this temperature,
it is clear that they reflect distinct phenomena, even without
extrapolation to locate $\phi_G$ and $\phi_J$ more precisely.
Correspondingly, the evolution of $\sigma_Y(\phi)$ 
in Fig.~\ref{zeroshear}b is influenced by both transitions, 
since solidity emerges near $\phi_G$, but $\sigma_Y$ increases 
significantly near $\phi_J > \phi_G$. This is consistent
with the idea that jamming mainly affects the 
very low temperature properties of the glass phase~\cite{hugo}. 

Finally, the rheology at $T=0$ corresponds to 
$\Pe = \infty$, and so the glass physics cannot operate.
Despite this complete change of regime, the 
corresponding flow curves shown in Fig.~\ref{flow}c
appear qualitatively very similar to the ones obtained at $T=10^{-4}$
in Fig.~\ref{flow}a. They are characterized by a 
Newtonian viscosity $\eta_0$ at small $\gammadot$ 
and low density, $\phi < \phi_J \approx 0.64$,
while a yield stress emerges upon compression, $\phi > \phi_J$.  
These data are fully equivalent to previous rheological studies of 
the athermal jamming transition~\cite{olson}, and indeed
near that transition can be collapsed using the same critical scaling.
The qualitative similarity between flow curves in Figs.~\ref{flow}a
and \ref{flow}c has created confusion in the literature~\cite{thomas},
where data obtained for systems undergoing the glass transition 
have been incorrectly analyzed in the athermal scenario of \cite{olson}.  

The shear viscosities $\eta_T$ and $\eta_0$ are measured
over $\gammadot$ windows that become well-separated 
at low $T$ and Fig.~\ref{zeroshear}a emphasizes that the difference
between the two functions increases as $T$ decreases, 
ruling out a smooth convergence of 
$\eta_T$ to $\eta_0$ for $T \to 0$. Instead, we 
find that as $T \to 0$, $\eta_T(\phi)$ follows the same density dependence as 
the equilibrium relaxation time of the corresponding 
hard sphere fluid~\cite{gio}, while $\eta_0$ is well described 
by an algebraic divergence~\cite{pouliquen}. 
Our results establish that the functions 
$G(\phi)$ and $J(\phi)$ 
controlling Newtonian flow in the hard sphere limit
are conceptually and quantitatively distinct. 

The yield stress is another highly sensitive indicator 
of the differences between glass and jamming transitions,
see Fig.~\ref{zeroshear}b. At finite $T$, solid behaviour 
emerges near $\phi_G \approx 0.59\ldots 0.61$, which agrees with equilibrium
dynamics studies~\cite{tom}. The yield stress then increases 
smoothly with $\phi$ up to $\phi_J \approx 0.64$ where its
density dependence changes dramatically. Also, while
$\sigma_Y$ scales with $T$ below $\phi_J$, it is of order unity (in
our units of $\epsilon/a^3$) above, with only a 
weak $T$-dependence scaling approximately as 
$\sim (\phi - \phi_J)$. Consistent with this picture, more detailed
analysis shows that $\sigma_Y(\phi,T)$
follows scaling behaviour near $\phi_J$ very similar to the one 
derived in Ref.~\cite{hugo2} for the pressure~\footnote{While shear stress 
and pressure closely match each other in solid phases, they differ 
in fluid phases since $P/T$ remains close to its equilibrium 
value for sheared thermal ($T>0$) 
fluids while $P \sim \gammadot$ at $T=0$.}.   
Thus glass and jammed states, having distinct 
physical origins, also display distinct stress scales,
and remain well-separated even as $T \to 0$ 
in the `glass-jamming phase diagram' shown in Fig.~\ref{zeroshear}c.
Note also that while the glass transition occurs at finite $T$ 
in the unsheared system, the jamming transition exists at $T=0$ only, 
so that these two limits never coexist.

The complex flow curves shown in Fig.~\ref{flow} can be 
modelled by assuming that the stress is an 
additive combination, $\sigma(\gammadot) = \sigma_G + \sigma_J + \eta_s 
\gammadot$, 
of contributions from glass and jamming 
physics, and from the solvent. 
The simplest model for the glass contribution 
incorporating the appropriate time and stress scales is 
\be
\frac{\sigma_G}{(k_B T/a^3)} =  \frac{\tilde{\Sigma}_G (\phi) }{  
[1+(\gammadot \tau_D)^\beta]^{-1}
+ [ \gammadot \tau_D \tilde\tau_G(\phi) ]^{-1} }, 
\label{sigmag}
\ee
with the dimensionless stress $\tilde{\Sigma}_G (\phi) =  \mbox{const.} 
+ (\phi-\phi_G)^\alpha (\phi_J - \phi)^{-\delta}
\Theta(\phi-\phi_G) \Theta(\phi_J - \phi)$, 
and a dimensionless timescale, $\tilde\tau_G(\phi)$,
diverging at $\phi_G$, e.g.\ as $\tilde\tau_G(\phi) \sim 
(\phi_G-\phi)^{-\gamma}$. In this model, 
the viscosity diverges at $\phi_G$, $\eta_T/\eta_s
\propto \tilde\tau_G(\phi)$ (with $\eta_s=\xi/a$), 
and a finite yield stress
emerges above, $\sigma_Y(\phi \ge \phi_G) = (k_B T/a^3)
\tilde{\Sigma}_G(\phi)$ which diverges near $\phi_J$. 
For the jamming contribution, we use a model consistent with the
scaling discussed in \cite{olson}, 
\be
\label{sigmaj}
\sigma_J / (\epsilon/a^3) = \tilde{\Sigma}_J(\phi) + 
[ (\gammadot \tau_0)^{-\beta'} + (\gammadot
\tau_0 J(\phi) )^{-1} ]^{-1}, 
\ee
where we take $\tilde{\Sigma}_J(\phi) \sim
(\phi-\phi_J)^{\alpha'}\Theta(\phi-\phi_J)$
and $J(\phi) \sim (\phi_J-\phi)^{-\gamma'}$. 
The entire set of data shown in Fig.~\ref{flow} can be
reproduced nearly quantitatively 
using exponents consistent with 
existing results: $\alpha = 0.7$~\cite{fuchs}, $\beta= 0.3$~\cite{BB},
$\delta = 0.8$~\cite{hugo2}
$\gamma=2.2$~\cite{tom}, 
$\beta'=0.4$, $\alpha'=1.2$ and $\gamma'=2.0$~\cite{olson}.  
Our empirical model, Eqs.~(\ref{sigmag}, \ref{sigmaj}), 
emphasizes that glass and jamming physics
take place over distinct time and stress sectors, and 
can be addressed by independent theoretical means. 
Differing predictions for, e.g., whether the onset of yield
stress~\cite{thomas,BBK} is continuous, as in theories of driven athermal
systems~\cite{sgr,hl}, 
or discontinuous as for driven glasses~\cite{fuchs} then make
physical sense.

\begin{figure}
\includegraphics[width=1.0\columnwidth]{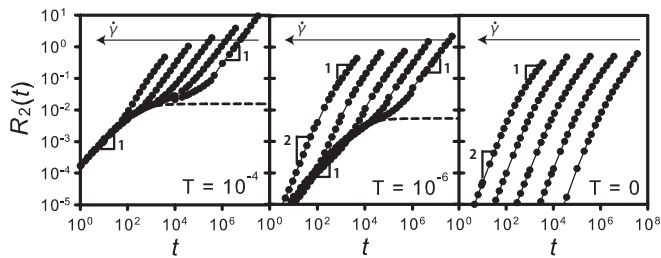}
\caption{
Mean-square displacement at $\varphi=0.62$  and 
$\dot{\gamma}=10^{-3} \cdots 10^{-7}$ and different temperatures. 
Caged dynamics (dashed) is only observed at finite $T$ and 
$\Pe \ll 1$.}
\label{msd}
\end{figure}

The distinct nature of thermal and 
athermal regimes is apparent also in microscopic dynamic correlation functions.
We plot the
mean-squared displacements, $R_2(t) = \langle | {\vec r}_i(t) - 
{\vec r}_i(0) |^2
\rangle$ in Fig.~\ref{msd} for fixed 
$\phi=0.62$ and different $T$ and $\gammadot$.
In the glass regime, $T=10^{-4}$, $R_2(t)$ displays 
short-time diffusion, caged dynamics at intermediate times, 
and shear driven diffusion at long times~\cite{yamamoto,BB}.
At $T=0$, we obtain very different, superdiffusive and diffusive,
behaviours, as discussed 
in Ref.~\cite{claus}. For intermediate 
temperature, $T=10^{-6}$, Fig.~\ref{msd} shows a clear crossover 
between thermal and athermal regimes: while caged dynamics is observed 
for low $\Pe$, superdiffusive motion is obtained at large 
$\Pe$. Therefore, the thermal-athermal crossover
observed in the macroscopic rheology in Fig.~\ref{flow}
originates from a similar crossover at the microscopic level.
While macroscopic flow curves in Fig.~\ref{flow}a-c can easily be confused, 
microscopic observations as in Fig.~\ref{msd} provide a 
clear qualitative distinction.

We have used 
temperature to study the crossover between two limits, while
experimentalists might equivalently tune particle softness.
We note from Figs.~\ref{flow}a and \ref{zeroshear}b
that for temperatures above $T \sim 10^{-5}$, corresponding to
thermal particle compressions $(1-r/a)\sim T^{1/2}$ of only
$10^{-3}\ldots 10^{-2}$,    
the $T=0$ physics has little influence on the rheological
behaviour.
This suggests that the jamming transition cannot be studied
using soft colloids unless $T/\epsilon$ is extremely small.
A second relevant experimental parameter is the particle size
setting the timescale for Brownian motion, with $\tau_D \sim 1$s
for particles of 1$\mu$m. This implies that the present thermal-athermal
crossover should be observable in experiments by tuning the particle size 
in the range 1-10$\mu$m. Yield stress data for emulsions~\cite{bibette} 
seem consistent with the data shown in Fig.~\ref{zeroshear}b,
but further studies are needed to confirm our predictions.

In conclusion, we have used shear rheology to study the relationship
between glass and
jamming transitions. While both correspond to the emergence of solid
behaviour as signalled by a finite yield stress, we have demonstrated
that they occur over stress and
time windows that become well-separated at low temperatures in dense
repulsive systems.  The glass-jamming phase diagram (Fig.~\ref{zeroshear}c)
has a scale-separated `wing' between $\phi_G$ and $\phi_J$, so that the
glass transition line does not extrapolate to the jamming point for $T \to
0$. The two transitions can only be observed separately in these two
distinct limits, i.e. on the glass line and at the jamming point. Any
other state point in the phase diagram is in principle affected by a
combination of both phenomena, in a way described by the simple
theoretical model of Eqs.~(\ref{sigmag}, \ref{sigmaj}). This conceptual
clarification should
help rationalize both experimental data and the scope of different
theoretical approaches.

\acknowledgments
We thank M. Pica Ciamarra for discussions, and 
R\'egion Languedoc-Roussillon for financial support (A.I., L.B.).

\end{document}